\documentclass[twocolumn,showpacs,amsmath,amssymb,superscriptaddress]{revtex4}
\usepackage{graphicx}
\usepackage{amsmath}
\usepackage{amssymb}
\usepackage{amsfonts}
\usepackage{epsfig}
\usepackage{color}

\newcommand{\ct}{\cite} 
\newcommand{\bi}{\bibitem}
\newcommand{\be}{\begin{equation}}
\newcommand{\ee}{\end{equation}}
\newcommand{\ba}{\begin{eqnarray}}
\newcommand{\ea}{\end{eqnarray}}

\newcommand{\non}{\nonumber}

\newcommand{\rf}[1]{(\ref{#1})}

\begin{document}
\title{Speeding up and slowing down the relaxation of a qubit by optimal control }

  \author{Victor Mukherjee}
 \affiliation{NEST, Scuola Normale Superiore and Istituto di Nanoscienze-CNR, I-56127 Pisa, Italy}
 \author{Alberto Carlini}
 \affiliation{NEST, Scuola Normale Superiore and Istituto di Nanoscienze-CNR, I-56127 Pisa, Italy}
\author{Andrea Mari}
 \affiliation{NEST, Scuola Normale Superiore and Istituto di Nanoscienze-CNR, I-56127 Pisa, Italy}

  \author{Tommaso Caneva}
 \affiliation{The Institute for Photonic Sciences, Mediterranean Technology Park, 08860 Castelldefels, Barcelona, Spain}

 \author{Simone Montangero}
 \affiliation{Institut fu\"r Quanteninformationsverarbeitung, Universita\"t Ulm, D-89069 Ulm, Germany}  
 
    \author{Tommaso Calarco}
 \affiliation{Institut fu\"r Quanteninformationsverarbeitung, Universita\"t Ulm, D-89069 Ulm, Germany}

   \author{Rosario Fazio}
 \affiliation{NEST, Scuola Normale Superiore and Istituto di Nanoscienze-CNR, I-56127 Pisa, Italy}
\author{Vittorio Giovannetti}
 \affiliation{NEST, Scuola Normale Superiore and Istituto di Nanoscienze-CNR, I-56127 Pisa, Italy}
 
\begin{abstract}

\centerline{ABSTRACT}
We consider a two-level quantum system prepared in an arbitrary initial state and relaxing to a steady state due to the 
action of a Markovian dissipative channel. We study how optimal control can be used for  speeding up or slowing down the relaxation towards the fixed point of the dynamics. We analytically derive the optimal relaxation times for different quantum channels in the ideal ansatz of unconstrained quantum control (a magnetic field of infinite strength). We also analyze  the situation in which the control Hamiltonian is bounded by a finite threshold. As byproducts of our analysis we find that: (i) if the qubit is initially in a thermal state hotter than the environmental bath, quantum control cannot speed up its natural cooling rate; (ii) if the qubit is initially in a thermal state colder than the bath, it can reach the fixed point of the dynamics in finite time if a strong control field is applied; (iii) in the presence of unconstrained quantum control it is possible to keep the evolved state  indefinitely and arbitrarily close to special initial states which are far away from the fixed points of the dynamics. 
 \end{abstract}  


\maketitle

\section{Introduction}

If a quantum system is not perfectly isolated from the environment it is subject to dissipation and decoherence and its dynamics is often well approximated by a Markovian quantum channel \cite{breuer02, huelga12}. In this case a given arbitrary initial state will usually converge towards a steady state and this process is called {\it relaxation}. The steady state can be the thermal state if the bath is in equilibrium; more generally, it will be a fixed point of the quantum channel describing the non-unitary evolution of the system. Depending on the situation, such a relaxation process can be advantageous or disadvantageous. If, for example, we want to cool a system by placing it into a refrigerator (or if we want to initialize a qubit), a fast thermalization is desirable. On the other hand, especially in quantum computation or communication, decoherence during processing is a detrimental effect and in this case a slow relaxation is preferable. The goal of this paper is to investigate how quantum control can be used to increase or decrease the relaxation time of a qubit towards a fixed point of the dynamics. 
The theory of optimal quantum control is well established and has been studied in a large variety of settings and under different perspectives (for a recent review see, e.g., \ct{brif}).
For example, the application of optimal control to open systems is discussed in \ct{tannor99} (cooling of molecular rotations), \cite{lloydviola} (using measurement), \ct{stefanatos04} (in the context of NMR), \ct{sklarz,jirari} (in $N$-level systems), \ct{rebentrost,hwang} (non-Markovian dynamics), \ct{roloff} (for a review).
In particular, time-optimal quantum control has been extensively discussed for one qubit systems in a dissipative environment [12-20],
a variational principle for constrained Hamiltonians in open systems can be found in \ct{carlini06,carlini08}, while a comparison of several numerical algorithms is given in \ct{machnes11}.
On the other hand, studies in closed \ct{margolus98} as well as open quantum systems \ct{campo13} pointed to the existence of upper bounds in the speed with which a quantum system can evolve in the Hilbert space (the `quantum speed limit', or QSL), and several applications of quantum control theory to achieve the QSL can be found in \ct{caneva09}.
An analysis of sideband cooling is given in \ct{wang11,rahmani13}, while
superfast cooling with laser schemes has proven to be advantageous \ct{machnes10}. 
More recently, the engineering of multipartite entangled quantum states via a quasilocal Markovian quantum dynamics has also been studied depending upon the available local Hamiltonian controls and 
dissipative channels (see, e.g., \cite{ticozzi} and references therein).
Time optimal quantum control has also been successfully applied in quantum thermodynamics \cite{kosloff}, e.g. to describe the fast cooling of harmonic traps \cite{hoffmann} or to maximize the extraction of work \cite{salamon}.

This work provides both analytical and numerical results. In the case in which the strength of the optimal control is allowed to be arbitrarily large, we give analytical expressions for the minimum and maximum relaxation times of a qubit subject to three prototypical classes of dissipative channels: generalized amplitude damping, depolarization and phase damping. 
For the amplitude damping channel we also analytically derive the results in the limit of a weak control field, as well as numerically optimize the relaxation time for different strengths of the control field using the chopped random basis (CRAB) optimization algorithm \ct{doria11}. We find that 
for initial hot thermal states the optimal path is a straight line towards the fixed point. This implies that it is impossible to speed up the cooling process of a thermal qubit in a cold bath by optimal control. However, optimal control can be advantageous if we want to heat a thermal qubit in the presence of a hot bath. Furthermore, in the limit of infinitesimal strength $m$ of a generic control Hamiltonian, the minimum time taken by a qubit to reach its fixed point decreases linearly with $m$, with the slope depending on the explicit form of the control Hamiltonian.
We also consider a different optimization task: to determine the maximum time for which one can keep the state of a qubit inside a ball of radius $\epsilon$ centered around the initial state. We show that, even if dynamical decoupling cannot be applied because the bath is Markovian, there exist special states for which the dissipative dynamics can be stopped by optimal control.   
In deriving our results we assume that the unitary (represented by a Hamiltonian) and the dissipative (represented by Lindbladians) parts act separately in the master equation governing the time evolution of the qubit. The Hamiltonian driving the qubit in the Bloch sphere can be controlled, subject to some constraints, in order to achieve our desired optimization task. However, the Lindbladians appearing in the master equation are fixed, time independent and not affected by any change in the system Hamiltonian. This is a reasonable assumption in the limit of very small changes in the strength of the system Hamiltonian, as well as in the opposite limit of an infinitely strong system Hamiltonian when any unitary evolution takes place almost instantaneously, during which time we can neglect the non-unitary part. Furthermore, we do not allow any feedback in our quantum control. 
 
The paper is organized as follows: in Section II we review the master equation describing the dynamics of a general dissipative and Markovian process and apply it to the case of two-level quantum systems whose state is represented in the Bloch sphere. We also introduce the problem of controlled time optimal evolution up to an arbitrarily small distance from the target. In Section III we discuss in more details the generalized amplitude damping channel. In Section III.A we analytically study how optimal control can speed up the relaxation of a qubit. In particular, Section III.A.1 is devoted to the case of unconstrained coherent control, while Section III.A.2 is devoted to the case of controls with constrained amplitude (with analytical results in the limit of small magnetic fields, and numerical results for arbitrary control amplitudes).
Then, the situation in which the control slows down the relaxation is treated in Section III.B.
Section IV deals with similar analytical studies of optimal control in the depolarizing channel, while Section V is devoted to the analysis of the phase damping channel. 
Finally, we provide some discussion of the results in Section VI. The general expression for the speed of change of purity of a qubit is given in the Appendix.

\section{Controlling the Markovian dynamics of a qubit}

A general dissipative and Markovian process can be described by the time-local master equation~\cite{breuer02,huelga12}: 
\ba
\dot{\rho} = -i\left[H,\rho \right] + \mathcal{L}\left(\rho\right), 
\label{master}
\ea
where $\rho(t)$ is the density operator representing the quantum system and $\dot \rho := \partial \rho /\partial t$.
Having set $\hbar=1$ for convenience, the Hermitian operator $H(t)$ describes the Hamiltonian of the system, which drives the unitary part of the quantum 
evolution.
 The superoperator ${\mathcal{L}}(\rho(t))$ instead is the dissipator, which is responsible for the decoherent part of the quantum evolution, and which 
can be expressed in terms of a collection of (in general non Hermitian) operators $L_a$ (the Lindblad operators) as in: 
\ba
\mathcal{L}\left(\rho(t)\right) :=\sum_a\left[ L_a \rho L_a^{\dagger} - \frac{1}{2} ( L_a^{\dagger} L_a \rho +  \rho L_a^{\dagger} L_a ) \right]\;.
\label{dissipator}
\ea
For a two-level quantum system, a qubit, the representation~(\ref{dissipator}) 
 can always be defined in terms of no more than three Lindblad operators $L_{a}$ ($a = 1,2,3$), which, exploiting the gauge freedom 
 inherent to the master equation~(\ref{master}), 
 can be chosen to be traceless, i.e. 
\ba
L_a := \sqrt{\gamma_a}~{\bf{l}}_a\cdot \boldsymbol{\sigma}, 
\label{las}
\ea
with $\boldsymbol{\sigma}:=(\sigma_x, \sigma_y, \sigma_z)$ being the vector formed by the  Pauli matrices $\{\sigma_i, ~ i=x, y, z\}$.
In this expression  
 ${\bf{l}}_a :=( l_{ax},  l_{ay},  l_{az})^\top$ are (possibly complex) 3-dimenstional vectors, fulfilling the  orthonormalization condition 
${\bf{l}}_a\cdot {{\bf{l}}}_b^\ast=\delta_{ab}$, while the non-negative parameters $\gamma_a$ define the decoherence rates of the system.
Analogously, without loss of generality the Hamiltonian $H$ can be written as:
\ba
H(t):={\bf{h}}\cdot \boldsymbol{\sigma},
\label{ham}
\ea
with ${\bf{h}}(t)$ being a 3-dimensional real vector. 
Accordingly Eq.~(\ref{master}) reduces to the following differential equation: 
\ba
\dot {\bf{r}}= 2\left [{\bf{h}} \wedge {\bf{r}} +   \sum_a \gamma_a [\Re
 (({\bf{l}}_a\cdot {\bf{r}}){\bf{l}}_a^\ast ) 
- {\bf{r}} + i ({\bf{l}}_a\wedge {\bf{l}}_a^\ast ) ]  \right ],
\label{parmaster}
\ea
where  ${\bf{r}}(t):=  (r_x, r_y, r_z)^\top$ is the  3-dimensional, real vector that represents the qubit density matrix $\rho$ 
in the Bloch ball, i.e. 
\begin{align}
\rho(t)= \frac{1}{2}\left(I + {\bf{r}}\cdot {\boldsymbol{\sigma}}\right)
\label{rho_para}
\end{align}
($I$ being the identity operator). For future reference it is worth reminding that 
while the Hamiltonian $H$ only induces rotations of the Bloch vector ${\bf{r}}$,
 the action of ${\cal L}$ typically will  modify also its length $r=|{\bf{r}}|$, i.e. the purity $P:=\mbox{Tr}[\rho^2]=(1+r^2)/2$ of the associated state $\rho$.

The main aim of our work is to study the time optimal, open-loop,
coherent quantum control of the evolution of one qubit state under the action of the master equation (\ref{parmaster}).
The coherent (unitary) control is achieved via the effective  magnetic field ${\bf{h}}(t)$ of Eq.~(\ref{ham}).
On the contrary, we assume the dissipative part of the quantum evolution~(\ref{dissipator})
fixed and assigned.  
 We also exclude the possibility of performing measurements on the system 
  to update the quantum control during the evolution, 
 i.e.  no feedback is allowed (notice however that complete information on the initial state of the qubit
 $\rho(t=0):=\rho_i = \left(I + {\bf{r}}_i\cdot {\boldsymbol{\sigma}} \right)/2$
  is assumed). 

Within this theoretical framework 
 we analyze  how to evolve the system  towards a target state $\rho_f:= \left(I + {\bf{r}}_f\cdot {\boldsymbol{\sigma}} \right)/2$
 in the shortest possible time. 
Specifically we take as 
$\rho_f$  a fixed point of the dissipative part  of the  master equation, i.e. 
a state $\rho_{\mathrm{fp}}:=({I + {\bf{r}}_{\mathrm{fp}}\cdot {\boldsymbol{\sigma}}})/{2}$
fulfilling the condition ${\cal{L}}(\rho_{\mathrm{fp}}) =  0$, or:
\ba
 \sum_a \gamma_a [\Re
 (({\bf{l}}_a\cdot {\bf{r}}_{\mathrm{fp}}){\bf{l}}_a^\ast ) 
- {\bf{r}}_{\mathrm{fp}} + i ({\bf{l}}_a\wedge {\bf{l}}_a^\ast ) ]=  0.
\label{fixpt}
\ea
Equation~(\ref{fixpt})  identifies stationary solutions (i.e. $\dot\rho=0$) of the master equation Eq.~(\ref{parmaster}) when no Hamiltonian is present. 
They represent attractor points for the dissipative part of evolution, i.e. states where noise would typically drive the system.
By setting $\rho_f= \rho_{\mathrm{fp}}$ in our time-optimal analysis we are hence effectively aiming at speeding up relaxation processes that would naturally occur in the system even in the absence of
external control. 
In addressing this  issue we do not require perfect unit fidelity, i.e. we tolerate that the quantum state arrives within a  small 
distance from the target, fixed a priori. More precisely, given $\epsilon\in[0,1]$ we look for the minimum value of time $T_{\mathrm{fast}}$ which thanks to a proper choice of
$H(t)$   allows us to satisfy
the constraint: 
\begin{eqnarray}
2D[\rho(T_{\mathrm{fast}}), \rho_f]= |{\bf{r}}(T_{\mathrm{fast}})-{\bf{r}}_{\mathrm{fp}}|=\epsilon, 
\label{tracedcdt}
\end{eqnarray}
with $D(\rho, \rho') := \mathrm{Tr}|\rho-\rho'|/2$ being the trace distance between the quantum states $\rho$ and $\rho'$ ~\cite{NIELSEN}.

A second problem we address is the exact counterpart of the one detailed above: namely we focus on  keeping the system  in its initial state $\rho_i$  
(or at least in its proximity)  
for the longest possible time. In other words, we try to slow down the relaxation which is naturally induced by ${\cal L}$ through the action of the control  Hamiltonian~$H$.

\section{Generalized amplitude damping channel}

\begin{figure}[ht]
\begin{center}
\includegraphics[width=8.5cm,angle=0]{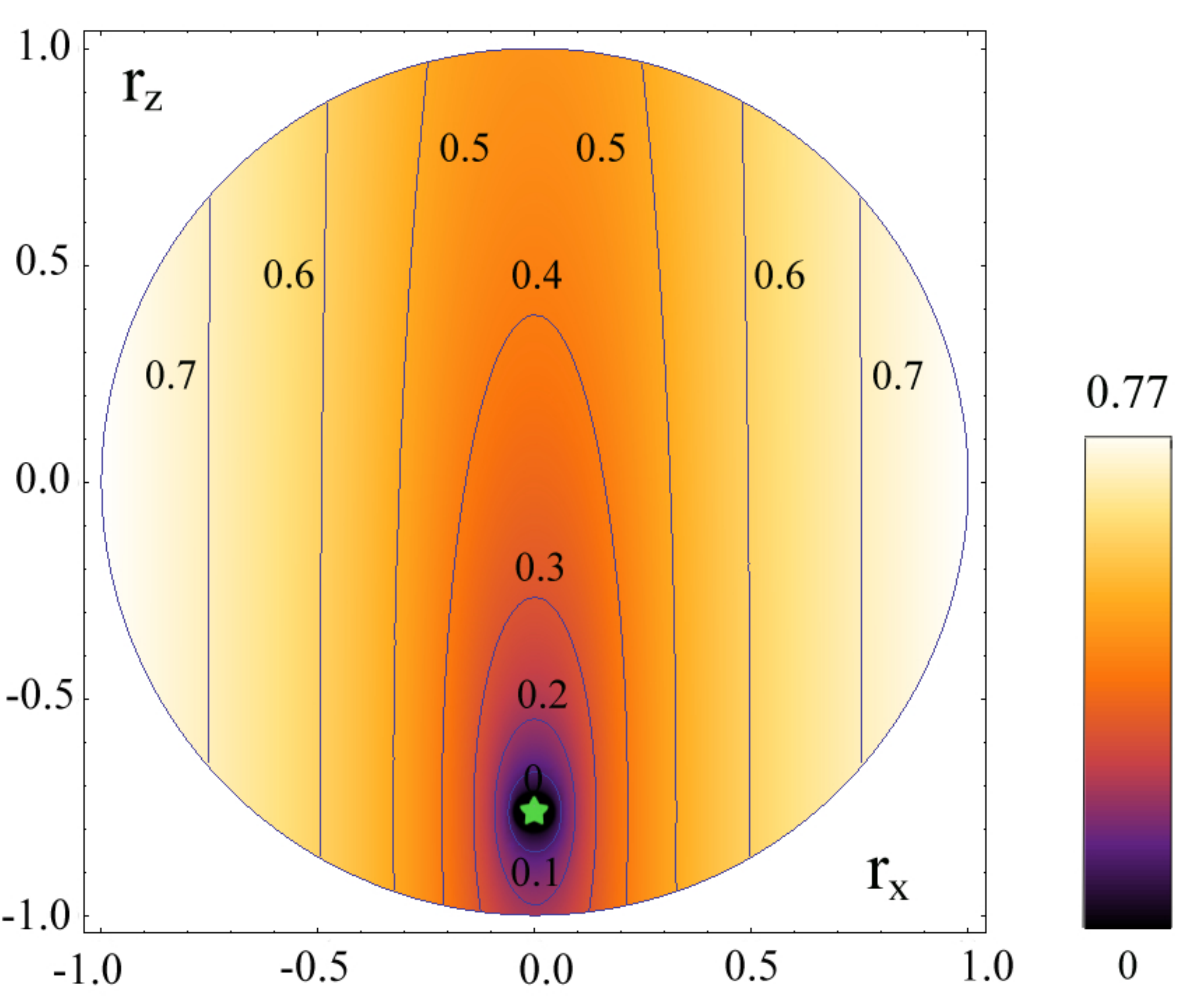}
\end{center}
\caption {Density plot of $T^{\mathrm{AD}}_{\mathrm{free}}({\bf{r}}_i,\epsilon)$ of Eq.~(\ref{adtime}) as a function of the initial state ${\bf{r}}_i = (r_{ix},r_{iy},r_{iz})$. As the system is invariant under rotations 
around the z-axis, we set $r_y=0$ without loss of generality. Here  $\epsilon = 0.04$ and the noise parameters have been set equal to $\beta=2$ and $\gamma= e^{\beta} -1\approx 6.39$. The fixed
point is indicated with a green star.}
\label{newfig}
\end{figure}

Here we analyze both the speeding up and the slowing down of relaxation  problems detailed in the previous section under the assumption that the dissipative dynamics~(\ref{dissipator})
 which is
affecting the system is 
a generalized amplitude damping channel~\cite{NIELSEN}. The latter is 
described by the Lindblad operators: 
\begin{eqnarray}
(L_1)_{AD}=\sqrt{\frac{\gamma}{e^\beta -1}} \sigma_+  ; ~~(L_2)_{AD}=\sqrt{\frac{\gamma e^\beta}{e^\beta -1}} \sigma_-,
\label{aplind}
\end{eqnarray}
where  $\sigma_\pm := (\sigma_x \pm  i \sigma_y )/2$, and where the non negative quantities  $\gamma$ and $\beta$ respectively describe
the decoherence rate of the system and the effective inverse temperature of the environmental bath.
In the absence of the Hamiltonian control, the associated superoperator ${\cal L}$ 
induces a dynamical evolution, which in the Cartesian coordinates representation~(\ref{parmaster}) is given by: 
\ba
\dot {\bf{r}}= -\frac{\gamma}{2r_{\mathrm{fp}}}(r_x, r_y, 2r_z)^\top -\gamma {(0,0,1)}^\top,
\label{admastereq}
\ea
with $r_{\mathrm{fp}}:=({e^\beta -1})/({e^\beta +1})$.
For an initial state  
${\bf{r}}_i:=(r_{ix}, r_{iy}, r_{iz})^\top$, Eq. \rf{admastereq} admits a solution of the form: 
\ba
{\bf{r}}(t)=e^{-\frac{\gamma t}{2r_{\mathrm{fp}}}}(r_{ix}, r_{iy}, e^{-\frac{\gamma t}{2r_{\mathrm{fp}}}}[r_{iz}+r_{\mathrm{fp}}] 
-e^{\frac{\gamma t}{2r_{\mathrm{fp}}}}r_{\mathrm{fp}})^\top , 
\label{admastersol}
\ea
which for sufficiently large  $t$ 
converges to the unique fixed point~(\ref{fixpt}) of the problem: 
\ba
{\bf{r}}_{\mathrm{fp}}=(0, 0, -r_{\mathrm{fp}})^\top \;.
\label{fixedad}
\ea 
From these expressions we can also compute the minimal time 
$T^{\mathrm{AD}}_{\mathrm{free}}({\bf{r}}_i,\epsilon)$ required for the initial state 
${\bf{r}}_i$  
 to reach the target ${\bf{r}}_{\mathrm{fp}}$ 
within a fixed trace distance $\epsilon$  without the aid of any external control, i.e. 
\ba
T^{\mathrm{AD}}_{\mathrm{free}}({\bf{r}}_i; \epsilon)&=&\frac{r_{\mathrm{fp}}}{\gamma}\ln\Biggl \{\frac{(r_{ix}^2+r_{iy}^2)}{2\epsilon^2} 
\nonumber \\
&\times &\left [1 +\sqrt{1+\left [\frac{2(r_{iz}+r_{\mathrm{fp}})\epsilon}{(r_{ix}^2+r_{iy}^2)}\right ]^2}\right ]\Biggr \}
\label{adtime}
\ea 
(see Fig.~\ref{newfig}). 
This function sets the benchmark that we  use to compare the performance of our time-optimal control problem.

\subsection{Speeding Up Relaxation} \label{sec:speedup}

In this section we address  the problem of speeding up the transition of the system from $\rho_i$ towards
the fixed point  state $\rho_{\mathrm{fp}}$ with a proper engineering of the quantum control Hamiltonian $H(t)$ to
see how much one can gain with respect to the  ``natural" time  $T^{\mathrm{AD}}_{\mathrm{free}}({\bf{r}}_i,\epsilon)$
of Eq.~(\ref{adtime}). Clearly the result will depend strongly on the freedom we have in choosing the functions ${\bf{h}}(t)$ of~Eq.~(\ref{ham}).

\subsubsection{Unconstrained Hamiltonian control} \label{sec:unc} 

\begin{figure}[ht]
\begin{center}
\includegraphics[width=8.5cm,angle=0]{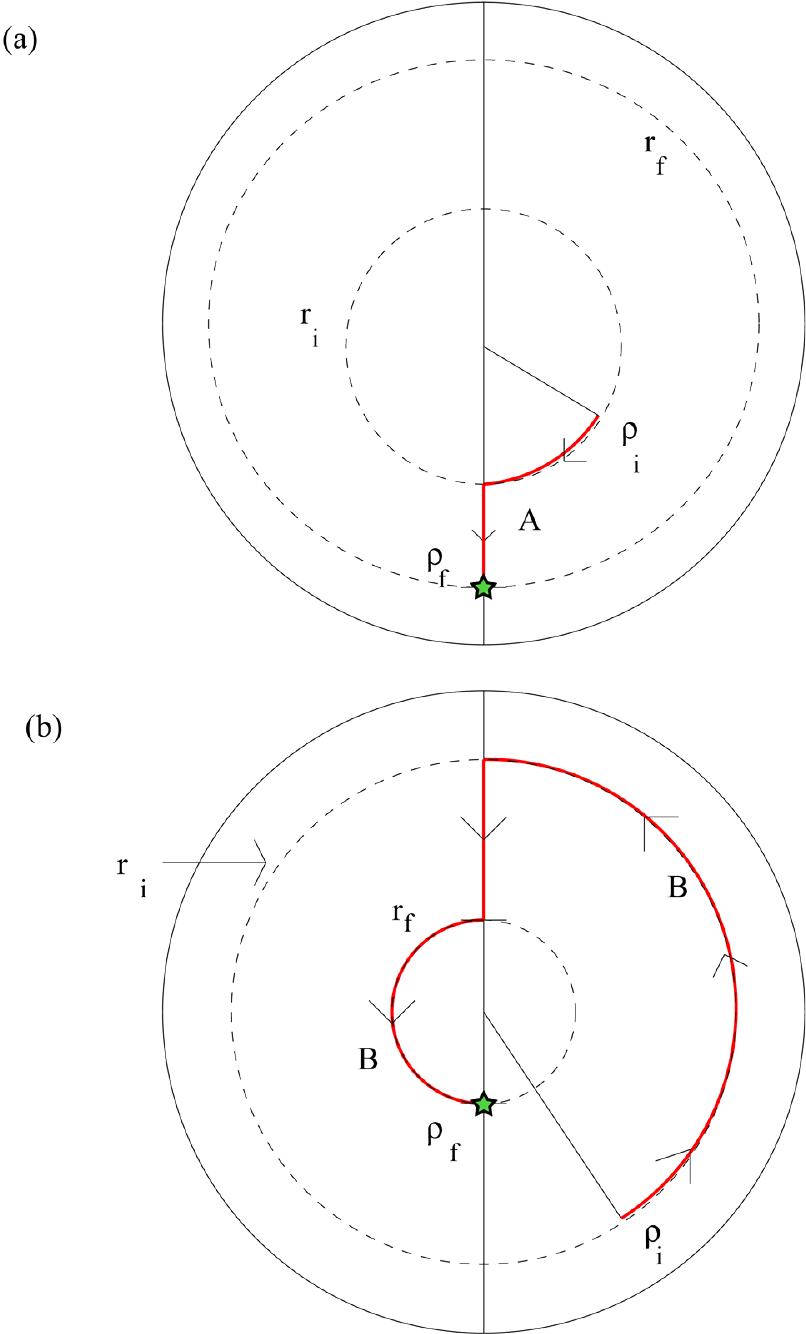}
\end{center}
\caption {Schematic diagram showing the optimal paths in the case of (a) cooling (path A) and (b) heating (path B) on the $x-z$ plane of the Bloch sphere. We start from an initial state $\rho_i$ with radius $r_i$. The fixed point  is given by $\rho_f$  with radius $r_f$ (green star). 
The solid vertical line is the $z$ axis.}
\label{bloch}
\end{figure}

For a coherent control where the choice  of the possible functions ${\bf{h}}(t)$ is unconstrained 
the problem  essentially reduces to finding the maximum of the modulus
of the speed of purity change, at any given purity, for the amplitude damping channel.
In fact, given any arbitrary initial state of the qubit (i.e., given an initial Bloch vector ${\bf{r}}_i$),
  one can always unitarily and instantaneously (since we may take a control with infinite strength) rotate the Bloch vector from the initial point along the surface of a sphere of radius $r_i$ until one reaches the new position of spherical coordinates $(r_i, \theta_{\mathrm{ext}}, \varphi_{\mathrm{ext}})$ where the speed of purity change induced by the dissipator, i.e. 
the quantity:
\ba
v[{\bf{r}}(P)]:= \frac{dP}{dt}=2 ~{\mathrm{Tr}}[\rho {\cal{L}}(\rho)],
\label{purity}
\ea
is extremal for fixed radius $r_i$. 
Then, one can switch off the control and let the system decohere for a time $T_{\mathrm{fast}}$ until the radius  $r(T_{\mathrm{fast}})$ 
which satisfies the trace distance condition (\ref{tracedcdt}) is reached.
Finally, one can switch the (magnetic field) quantum control on again and unitarily rotate the Bloch vector from
the position $(r(T_{\mathrm{fast}}), \theta_{\mathrm{ext}}, \varphi_{\mathrm{ext}})$ to a point within tolerable distance from the target at 
$(r(T_{\mathrm{fast}}), \theta_{\mathrm{fp}}, \varphi_{\mathrm{fp}})$.
Two examples of such a time optimal control  strategy are depicted  in Figs.~\ref{bloch} a-b, respectively for the cases $r_i < r_{\mathrm{fp}}$ and $r_i > r_{\mathrm{fp}}$.

From Eq.~(\ref{aplind}) and Eqs.~(\ref{generalspeed}), (\ref{generalspeedcoef}) of the Appendix
the speed of purity change in spherical coordinates induced by the generalized amplitude damping channel is easily shown to be independent of
the azimuthal angle $\varphi$ and given by: 
\begin{equation}
v_{\mathrm{AD}}(r, \theta)= -\gamma r \left [ \cos\theta + \frac{r}{2 r_{\mathrm{fp}}}(1+\cos^2\theta)\right ].
\label{v_ampl}
\end{equation}
The optimal values of the speed for a given radius $r$ are determined by the equation
$\partial_{\theta} v_{\mathrm{AD}}|_r=0$.
In the case of cooling, i.e. when we want to reach $r_{\mathrm{fp}}$ starting from 
$r_i  <  r_{\mathrm{fp}}$, we find that the speed $v_{\mathrm{AD}}$ is monotonically increasing from a negative minimum at $\theta_0=0$ (which corresponds to a global maximum of $|v_{\rm{AD}}|$) up to a positive maximum at $\theta_1=\pi$ (which corresponds to a local maximum of $|v_{\rm{AD}}|$).
Therefore the optimal cooling is achieved at $\theta_1=\pi$, where: 
\begin{equation} 
v^{\mathrm{AD, cool}}_{\mathrm{fast}}(r, \pi)=\gamma r \left (1-\frac{r}{r_{\mathrm{fp}}}\right ),     \quad r < r_{\mathrm{fp}}.
\label{vcool}
\end{equation}
Incidentally, this is consistent with the zero-temperature result considered in \cite{tannor99}.
On the other hand, in the heating case, i.e. when we want to reach the thermal state $r_{\mathrm{fp}}$ starting from 
$r_i  >  r_{\mathrm{fp}}$, 
the speed $v_{\mathrm{AD}}$ is always negative, it starts from a global minimum at $\theta_0=0$ (which again corresponds to a global maximum of $|v_{\rm{AD}}|$), grows up to a maximum at 
$\theta_2= \arccos (-r_{\mathrm{fp}}/r)$ (which corresponds to a global minimum of $|v_{\rm{AD}}|$) and then decreases to a local minimum at $\theta_1=\pi$ (which corresponds to a local maximum of $|v_{\rm{AD}}|$). 
Therefore, the optimal heating  is obtained by starting from $\theta_0 = 0$ where:
\begin{equation} 
v^{\mathrm{AD, heat}}_{\mathrm{fast}}(r, 0)=-\gamma r \left (1+\frac{r}{r_{\mathrm{fp}}} \right ),  \quad r > r_{\mathrm{fp}}.
\label{vheat}
\end{equation}
We remark here that, even if the above reasoning is valid in the regime of infinite strength of the control, nevertheless 
it gives also a no-go result for the task
of cooling a thermal hot state embedded in a cold bath. Since in this case the initial state is already along the negative z-axis, we cannot
increase the cooling time by optimal control and the fastest strategy is to just let the system thermalize with the bath.

We can finally proceed to compute the optimal time duration of the quantum controlled evolutions.
Using Eq. \rf{purity} and recalling the relationship between the purity and the Bloch vector of a given state, one can 
evaluate the required optimal time  from the optimal speeds Eqs.~(\ref{vcool}) and (\ref{vheat})  by the formula:
\ba
T^{\mathrm{AD}}_{\mathrm{fast}}({\bf r}_i; \epsilon) 
:=\left\{ \begin{array}{lll} 
\int_{r_i}^{r_{\mathrm{fp}}-\epsilon}\frac{rdr}{v^{\mathrm{AD,cool}}_{\mathrm{fast}}(r)}&& \mbox{for $r_i < r_{\mathrm{fp}}-\epsilon$} \\\\
0 && \mbox{for $|r_i - r_{\mathrm{fp}}| \leqslant \epsilon$} \\\\
\int_{r_i}^{r_{\mathrm{fp}}+\epsilon}\frac{rdr}{v^{\mathrm{AD,heat}}_{\mathrm{fast}}(r)}&& \mbox{for $r_i >  r_{\mathrm{fp}}+ \epsilon$,} 
\end{array} \right. \nonumber \\
\label{optime}
\ea
where we used $dP=rdr$. 

\begin{figure}[ht]
\begin{center}
\includegraphics[width=8.5cm,angle=0]{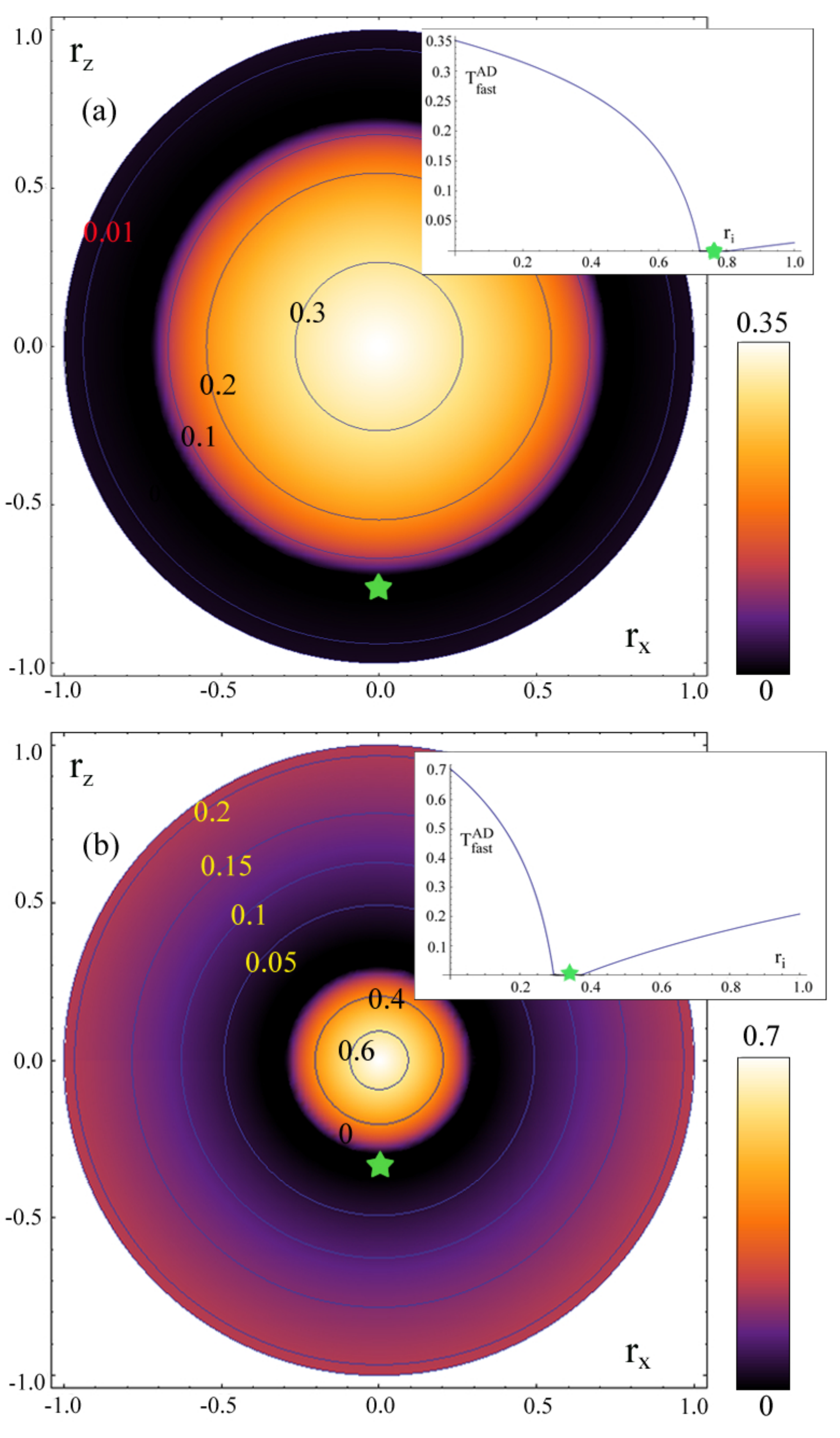}
\end{center}
\caption {Density plots of the minimal time $T^{\mathrm{AD}}_{\mathrm{fast}}({\bf r}_i; \epsilon)$ of Eq.~(\ref{optime}) for
a generalized amplitude damping channel as a function of the initial state ${\bf{r}}_i = (r_{ix},r_{iy},r_{iz})$. The parameters $\beta$ and $\epsilon$ are
as in Fig.~\ref{newfig} in (a), while $\beta =0.7$ and $\epsilon =0.04$ in (b). The fixed point is indicated by a green star.  The inset shows a section of the density plot along the x-axis.}
\label{newfig2}
\end{figure}

In particular, in the case of cooling, i.e. when we want to reach the target $r_{\mathrm{fp}}$ starting from $r_i  < r_{\mathrm{fp}}-\epsilon$, 
we obtain: 
\ba
T^{\mathrm{AD, cool}}_{\mathrm{fast}}({\bf r}_i; \epsilon) =\frac{r_{\mathrm{fp}}}{\gamma}\ln \left [\frac{(r_{\mathrm{fp}}-r_i)}{\epsilon}\right ], 
\label{adcoptime}
\ea
which, analogously to the free relaxation time~(\ref{adtime}),  diverges for $\epsilon \rightarrow 0$. 
In the case of heating, i.e. when we want to reach the target $r_{\mathrm{fp}}$ starting from $r _i > r_{\mathrm{fp}}+ \epsilon$ we obtain
\ba
T^{\mathrm{AD, heat}}_{\mathrm{fast}}({\bf r}_i; \epsilon) =\frac{r_{\mathrm{fp}}}{\gamma}\ln \left [\frac{(r_{\mathrm{fp}}+r_i)}{(2 r_{\mathrm{fp}}+\epsilon)}\right ].
\label{adhoptime}
\ea
This time is finite even in the limit of $\epsilon \rightarrow 0$, and it clearly represents an advantage with respect to the action of simply letting the system evolve without any control from the initial state (cf. Eq.  \rf{adtime} for $\epsilon \rightarrow 0$, also see Figs. \ref{newfig} and \ref{newfig2}).  

We notice finally that, to the most significant order in an expansion in $\epsilon$,  the function~(\ref{optime})  reaches its maximum for ${\bf{r}}_i=0$, i.e. 
\begin{eqnarray}
 \max_{{\bf r}_i} \; T^{\mathrm{AD}}_{\mathrm{fast}}({\bf r}_i; \epsilon)  \simeq \frac{r_{\mathrm{fp}}}{\gamma}|\ln \epsilon|.
 \end{eqnarray}
 This is the optimal time one would have to wait in the worst possible scenario (of choice of initial conditions) in order to bring the system close to the target in the case of unconstrained control.
 By comparing it with the maximum of the function (\ref{adtime}), i.e. 
 \ba
 \max_{{\bf r}_i} T^{\mathrm{AD}}_{\mathrm{free}}({\bf{r}}_i; \epsilon) \simeq 2 \frac{r_{\mathrm{fp}}}{\gamma}|\ln \epsilon|,
 \ea
(reached by a pure state along the equator of the Bloch ball) we notice that the optimal quantum control yields a shortening of a factor two in the evolution time. 

\subsubsection{Optimal control with constrained magnetic field intensity}

\begin{figure}[ht]
\begin{center}
\includegraphics[width=8.5cm,angle=0]{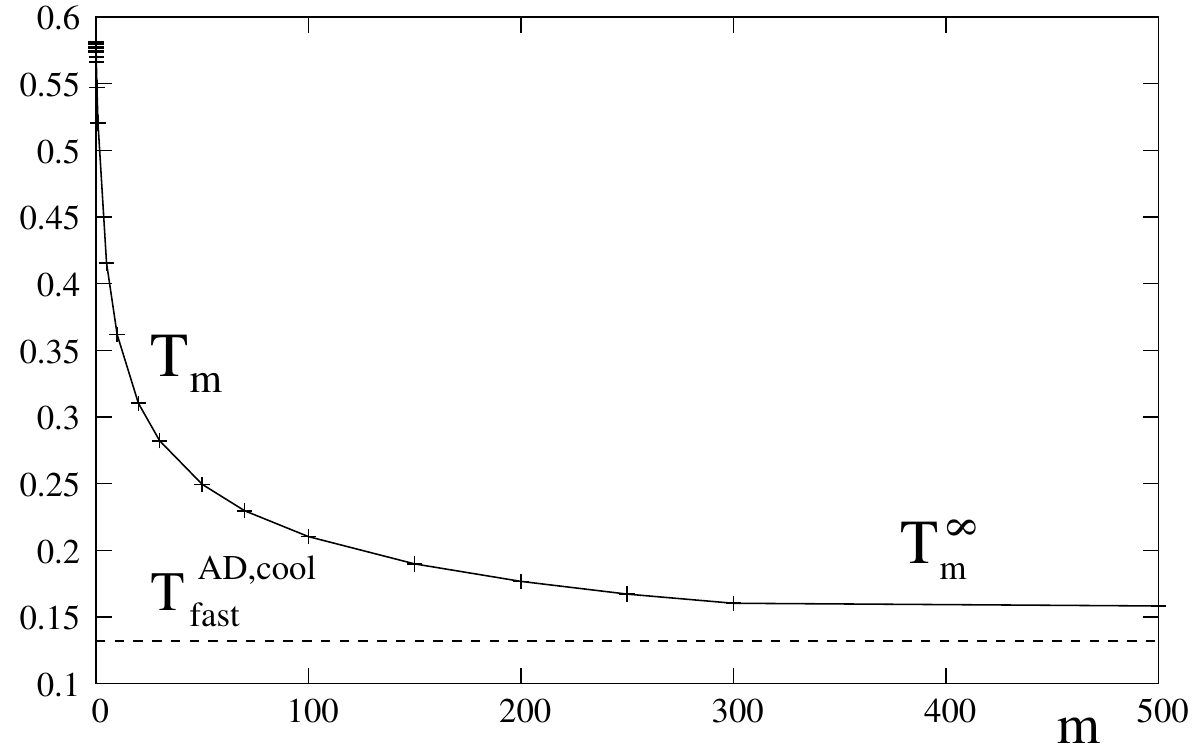}
\end{center}
\caption {Plot of  the optimal time evolution  $T_m$ needed to bring the initial state  $\rho_i$ with $\textbf{r}_i = (0.38, -0.22, -0.46)$ towards the fixed point of a generalized
amplitude damping channel with $\beta = 2$ and $\gamma = e^\beta-1$. Data obtained
via numerical optimization of the control parameters $h_{\mu,n}$ of Eq.~(\ref{crabb}) setting $\tau = N_c = 10$ and $\epsilon = 0.04$. 
In the limit of $m\rightarrow \infty$ and of $N_c\rightarrow\infty$ we expect $T_m$ to saturate to the corresponding value of the function 
$T^{\mathrm{AD, cool}}_{\mathrm{fast}}({\bf r}_i,\epsilon)$ given in  Eq.~(\ref{adcoptime}) (dashed line).}
\label{topt_m}
\end{figure}

The results of the previous section have been obtained under the assumption of an unconstrained Hamiltonian control.
Of course this is a highly idealized scenario which may not be approached  in realistic experimental setups.
On the contrary,  the effective  magnetic field ${\bf{h}}(t)$ entering in Eq.~(\ref{ham}) 
contains an uncontrollable, fixed part ${\bf{h}}^D(t)$  ({\em{drift}} contribution) which can be only in part 
compensated via the application of
some controlling pulse ${\bf{h}}^C(t)$ whose maximum  intensity is bounded by a fixed, finite  value $m$,~i.e. 
\begin{eqnarray} \label{newham}
{\bf{h}}(t):= {\bf{h}}^D(t) +  m \; {\bf{h}}^C(t),    \qquad |{\bf{h}}^C(t)|\leqslant 1. 
\end{eqnarray} 
Discussing the speeding up of relaxation under these conditions  is a rather complex task for which at present we do not have
an analytical solution (apart from the special case where $m$ is small, see below). Still, in the following we present a numerical analysis  that allows us to
gain some insight into the problem.
In particular, we focus on the case where the initial state of the system $\rho_i$ is
characterized by a Bloch vector of length  $r_i=0.41$ (specifically we take  
$\textbf{r}_i = (0.38, -0.22, -0.46)$
and take $\beta=2$ and $\gamma= e^{\beta}-1$ as 
parameters for the generalized amplitude damping channel. Accordingly, 
this corresponds to have Lindblad generators~(\ref{aplind})  equal to 
$(L_1)_{AD} =\sigma_+$, $(L_2)_{AD} = e \sigma_-$ and a fixed point  (\ref{fixedad}) with $r_{\mathrm{fp}}\simeq 0.76$. 
For the Hamiltonian~(\ref{newham}), moreover, we take:
\ba
{\bf{h}}^D(t) &=&  \frac{\omega}{2} \; {\bf{e}}_z + \frac{t}{\tau}\left({\bf{e}}_x+{\bf{e}}_y+{\bf{e}}_z \right), \label{crabb1}\\
{\bf{h}}^C(t) &=& \frac{t}{\tau N_c}\sum_{n=1}^{N_c}\sum_{\mu =x, y, z} h_{\mu,n} \sin\left (\frac{2\pi n t}{\tau}
\right ) {\bf{e}}_{\mu},
\label{crabb}
\ea
where $\{ {\bf{e}}_{\mu}, \mu =x,y,z\}$ are the Cartesian unit vectors. 
The control term ${\bf{h}}^C(t)$ is chosen following the methods of CRAB \ct{doria11}.
The drift term ${\bf{h}}^D(t)$ contains  two contributions: a constant term
which sets the energy scale for the qubit  and a time dependent term describing  side effects  of the control process 
(in particular we model it as an isotropic increase of the magnetic field over the duration time of the evolution).
The control pulses  to be optimized are finally represented  in terms of a  truncated Fourier expansion containing $N_c$ terms
whose coefficients  are subject to the constraints $-1 < h_{x,n},h_{y,n}, h_{z,n} < 1$, for  all $n$.
For a given value of the intensity bound $m$,  we then use  a simplex method \ct{doria11} to numerically optimize $h_{\mu,n}$ so that 
the system, starting from $\rho_i$, will get to  a (trace) distance  $\epsilon = 0.04$ from 
the fixed point $\rho_{\mathrm{fp}}$   in the shortest possible time $T_m$. 
Results are reported in Fig.~\ref{topt_m}: 
 as expected, $T_m$ decreases monotonically with $m$, converging to a constant value $T^\infty_m$ at large $m$. 
 As we are simulating a cooling process ($r_i$ being smaller than $r_{\mathrm{fp}}$)
 the latter should be compared with the analytic value of $T^{\mathrm{AD, cool}}_{\mathrm{fast}}({\bf r}_i,\epsilon)$ of
  Eq.~(\ref{adcoptime})  where an unbounded (both in the intensity $m$ and in the
  frequency domain) Hamiltonian control was explicitly assumed. The value of $T^{\mathrm{AD, cool}}_{\mathrm{fast}}({\bf r}_i,\epsilon)$
 is represented by the dashed line of Fig.~\ref{topt_m}: 
the discrepancy  between $T^\infty_m$ and  $T^{\mathrm{AD, cool}}_{\mathrm{fast}}({\bf r}_i,\epsilon)$ 
is expected to saturate in the limit of a large $m$ and number $N_c$ of frequencies in Eq. (\ref{crabb}).

\begin{figure}[ht]
\begin{center}
\includegraphics[width=8.5cm,angle=0]{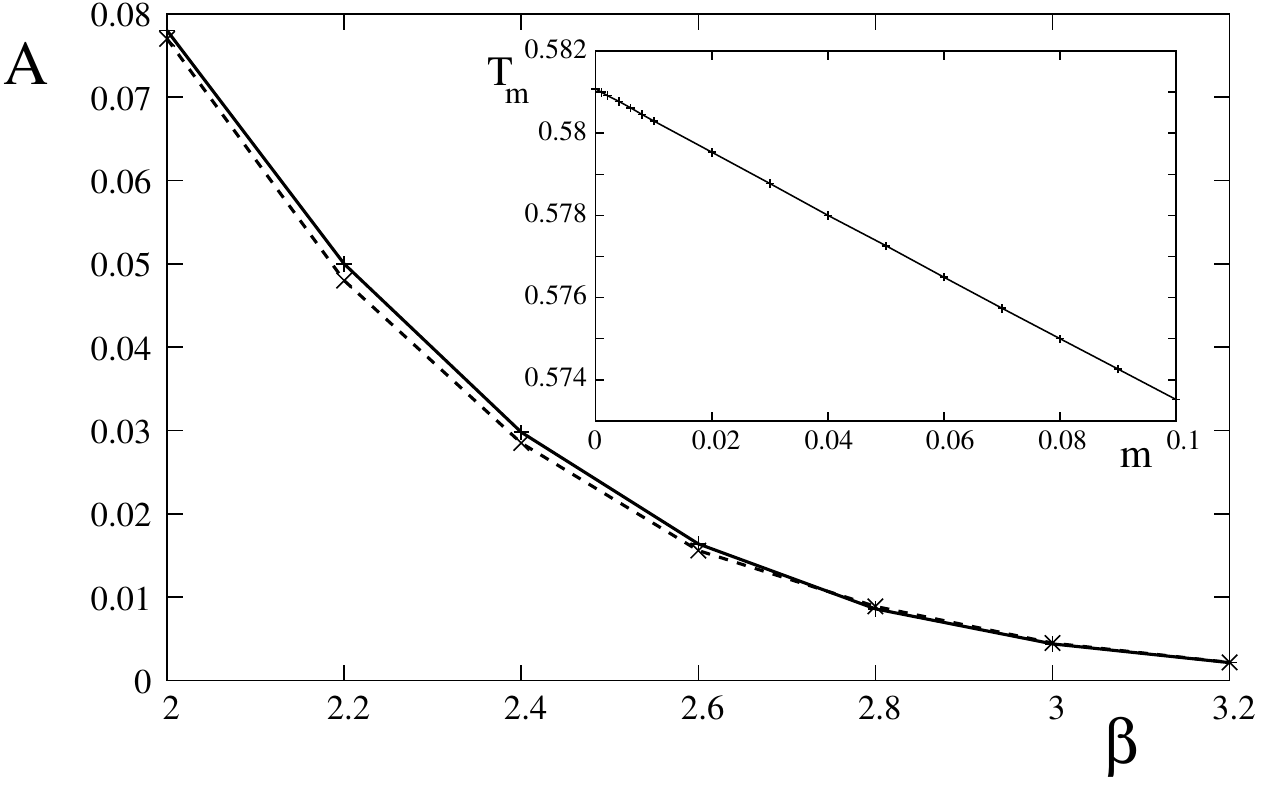}
\end{center}
\caption { Comparison between the numerical (solid line) and the analytical bound (\ref{anbound}) (dashed line) values of the slope $A$ as a function of $\beta$ for  
$N_c = \tau = 10$ and $\epsilon = 0.04$. The initial point ${\bf{r}}_i$ is the same as in Fig. 4. $m/\dot{\theta}_0$ decreases for larger values of $\beta$, thus resulting in a better match between the numerical and analytical values in this regime. 
Inset: variation of $T_m$ as a function of $m$ for small $m$ for $\tau = N_c = 10$, $\beta = 2$ and $\epsilon = 0.04$. As expected, $T_m$ decreases linearly with $m$. }
\label{A_beta}
\end{figure}

Now let us focus on the small $m$ limit.
 To do so we find it convenient to write 
the master equation (\ref{parmaster}) for the generalized  amplitude damping channel in terms of the spherical coordinates ($r(t), \theta(t), \varphi(t)$)  of 
the vector ${\bf{r}}(t)$, i.e. 
\ba
\dot{r} &=& -\frac{[r(1 + \cos^2\theta ) + 2 r_{\rm{fp}}\cos \theta]}{(1-r_{\rm{fp}})} 
\non\\
\dot{\theta} &=& \frac{\sin\theta(r\cos\theta +2 r_{\rm{fp}} )}{r(1-r_{\rm{fp}})}+2(-h_x \sin\varphi + h_y \cos\varphi )
\non \\
\dot{\varphi} &=& -2\left [ (h_x \cos\varphi + h_y\sin\varphi)\cot \theta  - h_z \right ],
\label{master_m}
\ea
where $h_{x}(t)$, $h_{y}(t)$,  and $h_{z}(t)$ are the Cartesian components of the Hamiltonian vector \rf{newham}.
For the moment let us consider the case $m= 0$ (no control). When $r < r_{\rm{fp}}$,  from  Eq. (\ref{master_m}) 
we have that $\dot{\theta} > \sin\theta(\cos\theta +2)/(1-r_{\rm{fp}}) > 0$ for any $t$, i.e. $\theta$ increases monotonically in the cooling case. On the other hand, in the case of heating, even though $r > r_{\rm{fp}}$ implies that $\dot\theta $ can be negative at small times (when the system is far away from the fixed point), at large times when $r \approx r_{\rm{fp}}$ we have $\dot{\theta} \approx \sin\theta(\cos\theta +2)/(1-r_{\rm{fp}})$, and thus again $\theta$ increases monotonically. 
These behaviors will be  maintained also for $m\neq 0$ as long as $m$ is sufficiently small.
Therefore, as $\theta$ is almost monotonic in time for all possible choices of the input state (the only exceptions being for heating processes), we can use it
 to parametrize the trajectories of the system.
 This allows us to write  the time $T_m$ taken by the qubit to move from the initial state to a state within trace distance $\epsilon$ of the fixed point 
 as:
\ba
T_m = \int^{\theta_m}_{\theta_i} \frac{d\theta}{\dot{\theta}} 
= \int^{\theta_m}_{\theta_i} \frac{d\theta}{\dot{\theta}_0 + m\; \Gamma},  \label{eqforgamma}
\ea
where $\Gamma (t(\theta)):= 2\left[-h^C_{x} \sin\varphi + h^C_{y}\cos\varphi \right]$, $\dot{\theta}_0(t(\theta)) := \dot{\theta}$ at $m=0$ and $(r_m, \theta_m, \varphi_m)$  (respectively $(\bar{r}, \bar{\theta}, \bar{\varphi})$) are the coordinates of the final state for $m\neq 0$ (respectively $m=0$). In the limit $m\Gamma \ll \dot{\theta}_0$ and expanding for small $m$ we get:
\ba
T_m \approx 
  \bar{T} - m A,
 \label{weakmtime}
\ea
where
\ba
\bar{T} := \int^{\bar{\theta}}_{\theta_i} \frac{d\theta}{\dot{\theta}_0}
\ea
is the time taken to reach the fixed point at $m=0$ and
\ba
A:= \int^{\bar{\theta}}_{\theta_i} \frac{\Gamma(\theta)}{\dot{\theta}^2_0} d\theta- \left[\frac{1}{\dot{\theta}_0}\right]_{\bar{\theta}} \left[\frac{\partial \theta_m}{\partial m}\right]_{m=0}.
\label{A}
\ea
 Assuming that $\dot{r} = \dot{\bar{r}}_0 := \dot{r}(t = \bar{T}, m=0)$ is a constant for $T_m \leq t \leq \bar{T}$, and using the trace distance 
criteria (\ref{tracedcdt}),
it can be shown that:
\ba
A= \frac{1}{(1-D)}\left[\int^{\bar{\theta}}_{\theta_i} \frac{\Gamma(\theta)}{\dot{\theta}_0^2}  d\theta\right] ,
\label{bound}
\ea
where $D = \dot{\bar{r}}_0\left(\bar{r} + r_{\rm{fp}}\cos\bar{\theta} \right)/(\dot{\bar{\theta}}_0\bar{r}r_{\rm{fp}}\sin \bar{\theta})$ and $\dot{\bar{\theta}}_0$ is $\dot{\theta}_0$ at $\theta = \bar{\theta}$.
Eqs. \rf{weakmtime} and \rf{A} clearly show that, in the limit in which the magnetic field used for quantum control has small amplitude, the optimal time to reach the target
fixed point within trace distance $\epsilon$ decreases linearly with $m$ for the qubit in the amplitude damping channel.
To validate the above analysis we have again adopted numerical techniques assuming a temporal dependence for ${\bf{h}}^C(t)$ as in Eq.~(\ref{crabb}) --
results are reported in Fig.~\ref{A_beta}.
In these simulations the value of ${h}_{\mu,n}$ is fixed at the beginning of an iteration and it cannot  change during the course of the evolution. Therefore $|h^c_{\mu}|$ can take its maximum possible value of $\alpha(t)= \frac{1}{N_c}\frac{t}{\tau} \sum_{n = 1}^{N_c} |\sin(2\pi n t/\tau)|$ only if $\sin \left(2\pi n t/\tau \right)$ has the same sign for any $t$ and for a particular $n$, i.e. $2\pi N_c T_m/\tau \leq \pi$. Again, from the definition of $\Gamma$ of Eq.~(\ref{eqforgamma}), we get $\Gamma \leq 2\left(|\sin\varphi| + |\cos\varphi| \right) \alpha$. Therefore using Eq. (\ref{bound}) we finally arrive at an upper bound for the slope $A$, given by:
\ba
A &\leq  &\frac{2}{N_c \tau(1-D)}\sum_{n = 1}^{N_c}
\nonumber \\
&\times & \int^{\bar{\theta}}_{\theta_i} \frac{t\left(|\sin\varphi| + |\cos\varphi| \right)}{\dot{\theta}_0^2}  \left|\sin \left [\frac{2\pi n t}{\tau} \right ]\right | d\theta .
\label{anbound}
\ea

\subsection{Slowing Down Relaxation}

Here we are interested in the opposite problem to that analyzed so far.
In other words, we would like to find out for how long a qubit subject to amplitude damping can be kept, with the aid of a quantum control represented by a magnetic field of infinite maximum strength, arbitrarily close to a given initial state ${\bf{r}}_i$.
Again, one can quantify the notion of closeness by imposing that the trace distance between the evolved state and the initial state is arbitrarily small.
In other words, we are interested in applying the optimal control such that  $|{\bf{r}}_i - {\bf{r}}(t)| \leq \epsilon$ for the maximum time duration $T^{\rm{AD}}_{\rm{slow}}$. 
On the one hand, we are free to control the Bloch vector of the qubit unitarily and instantaneously in the directions tangent to the sphere of radius $r_i$.
On the other hand, the qubit will be subject to uncontrollable decoherence along the radial direction, with its purity changing at speed $v$.
Here we confine ourselves to the explicit analysis of the case in which the relaxation dynamics can be controlled for an indefinitely long time
\footnote{We note that one could define the problem in other ways, namely one could allow for the quantum state to evolve along a trajectory which crosses the $\epsilon$-ball around ${\bf{r}}_i$ several times before finally returning inside it, and calculate the maximal time for which this dynamics is possible.}.

For the amplitude damping channel, in the case of an initial state with $r_i  < r_{\mathrm{fp}}$ we can see that the speed $v_{\mathrm{AD}}$  (and equivalently $\dot{r}(t)$) becomes zero as we approach the angle (see Fig. \rf{contour})
\ba
\theta_3:= \arccos \left [\frac{r_{\mathrm{fp}}}{r_i}\left ( \sqrt{1-\frac{r_i^2}{r^2_{\mathrm{fp}}}} -1 \right )\right ].
\label{adcanglemax}
\ea
Thus, if the quantum state of the qubit happens to have initial polar angle $\theta_3$, quantum control with infinite strength will be able to keep the qubit there indefinitely, i.e., $T^{\rm{AD}}_{\rm{slow}} \to \infty$ for these initial states.
This is because for any point along the ellipsoid defined by Eq. \rf{adcanglemax} the velocity $\dot{\bf{r}}$ is orthogonal to the Bloch vector and therefore it can be controlled by unitaries.
In a sense, one could say that unbounded coherent control has allowed to extend the set of fixed points by adding the set of points with $v=0$.

\begin{figure}[ht]
\begin{center}
\includegraphics[width=8.5cm,angle=0]{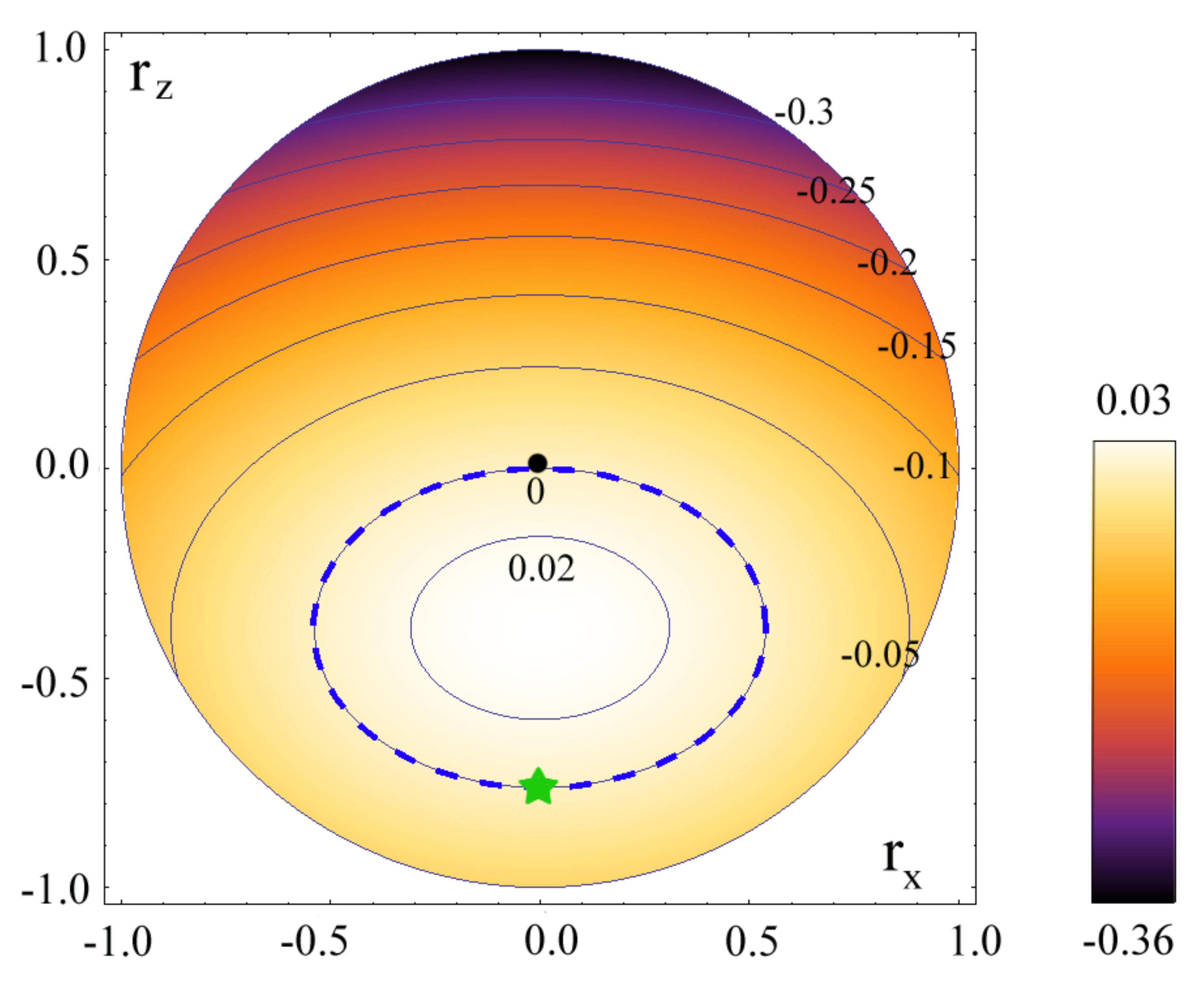}
\end{center}
\caption {Plot of the speed $v_{\mathrm{AD}}$  on the $x-z$ plane of the Bloch sphere when the fixed point is the thermal state corresponding to $\beta = 2$. The curve $v_{\mathrm{AD}} = 0$ (dashed line) is an ellipse which passes through the origin of coordinates (black dot) and the fixed point (green star).}
\label{contour}
\end{figure}

\section{Depolarizing Channel}

In this section we address the problem of quantum control of the relaxation when 
the dissipative process affecting the system is a depolarizing channel~\cite{NIELSEN}. 
The latter  is characterized by the three Lindblad operators
\ba
&(L_1)_{DP}=\sqrt{{\gamma_x}} \sigma_x ; ~(L_2)_{DP}=\sqrt{{\gamma_y}} \sigma_y ;& 
\nonumber \\
&(L_3)_{DP}=\sqrt{{\gamma_z}} \sigma_z.&
\label{dplind}
\ea
and it admits as unique fixed point the fully mixed state $\rho_{\mathrm{fp}}=I/2$, i.e. ${\bf{r}}_{\mathrm{fp}}=0$. 
In the absence of unitary control, the associated master equation~(\ref{parmaster}) is given by: 
\ba
\dot {\bf{r}}= -2(\Gamma_x r_x, \Gamma_y r_y, \Gamma_z r_z)^\top,
\label{adpmastereq}
\ea
where $\Gamma_x:=\gamma_y+\gamma_z$, $\Gamma_y:=\gamma_x+\gamma_z$, $\Gamma_z:=\gamma_x+\gamma_y$, 
with solution, for the initial condition ${\bf{r}}_i:=(r_{ix}, r_{iy}, r_{iz})^\top$,
\ba
{\bf{r}}(t)= (e^{-2\Gamma_x t}r_{ix}, e^{-2\Gamma_y t}r_{iy}, e^{-2\Gamma_z t}r_{iz})^\top.
\label{adpmastersol}
\ea
The relaxation time $T^{\mathrm{DP}}_{\mathrm{free}}({\bf r}_i; \epsilon) $
from an arbitrary initial state $r_i$ to the fixed point in the absence of quantum control can be found from the trace distance condition (\ref{tracedcdt}) and from the solution \rf{adpmastersol} 
by solving the implicit equation:
\ba
|{\bf{r}}[T^{\mathrm{DP}}_{\mathrm{free}}({\bf{r}}_i; \epsilon)]| =\epsilon.
\label{adptime}
\ea

Moreover, from Eq.~(\ref{dplind}) and Eqs.~(\ref{generalspeed}), (\ref{generalspeedcoef}) of the Appendix, 
the speed of purity change in spherical coordinates reads:
\ba
v_{\mathrm{DP}}(r, \theta, \varphi)&=& -r^2 \{2\Gamma_z +[(\Gamma_x +\Gamma_y -2\Gamma_z) 
\non \\
&+& (\Gamma_x-\Gamma_y) \cos2\varphi ] \sin^2\theta \}.
\label{vdep}
\ea
This velocity is always negative and it is easy to check that  its absolute value is maximum at the intersection 
of the sphere of radius $r$ with the coordinate axis associated with the minimum value among $\gamma_x, \gamma_y$ and $\gamma_z$.  The optimal heating velocity is then
\begin{equation} 
v^{\mathrm{DP, heat}}_{\mathrm{fast}}(r)=-2 \Gamma_M r^2, 
\label{vphaseheat}
\end{equation}
where $\Gamma_M$ is the largest among  $\Gamma_x, \Gamma_y$ and $\Gamma_z$. 
Note that, in the special case when any two of the decay rates are equal, one has families of optimal solutions along the circle that is the intersection between the sphere of radius $r$ and the plane of coordinates corresponding to the equal decay rates. 
Moreover, in the completely symmetric case 
of $\gamma_x=\gamma_y=\gamma_z :=\gamma_0$ the heating speed is given by
$v^{\mathrm{DP, heat}}_{\mathrm{fast}}=-4 \gamma_0 r^2$ 
for all angles $\theta$ and $\varphi$.  Therefore in this case any control is useless. 

Inserting the maximal speed (\ref{vphaseheat}) into Eq.~(\ref{optime}) we obtain the optimal time:
\ba
T^{\mathrm{DP, heat}}_{\mathrm{fast}}({\bf r}_i; \epsilon) =\frac{1}{2\Gamma_M}\ln \left [\frac{r_i}{\epsilon}\right ].
\label{adpoptime}
\ea

The function~(\ref{adpoptime})  reaches its maximum for a pure state along one of the coordinate axis, i.e. 
\begin{eqnarray}
 \max_{{\bf r}_i} \; T^{\mathrm{DP}}_{\mathrm{fast}}({\bf r}_i; \epsilon)  =\frac{|\ln\epsilon|}{2\Gamma_M}.
 \end{eqnarray}
 This is the largest time one would need to wait in order to bring the system close to the target in the case of unconstrained control.
 By comparing it with the maximum for the free relaxation time obtainable from Eq. \rf{adptime}, i.e. 
 \ba
 \max_{{\bf r}_i} T^{\mathrm{DP}}_{\mathrm{free}}({\bf{r}}_i; \epsilon) =\frac{|\ln\epsilon|}{2\Gamma_m},
 \ea
where $\Gamma_m$ is the smallest among the $\Gamma_x, \Gamma_y$ and $\Gamma_z$ (reached for a pure state along one of the axis) we notice that the optimal time control yields a shortening by a factor $\Gamma_m/\Gamma_M$ in the evolution time.

In this case the set of points with $v_{\mathrm{DP}}=0$ coincides with the set of fixed points and, therefore, any control is useless for stopping the relaxation.

\section{Phase Damping Channel}

The phase damping channel 
is a dissipative process characterized by a single Lindblad operator:
\begin{eqnarray}
(L_1)_{PD}&=&\sqrt{{\hat \gamma}} \sigma_z,
\label{pdlind}
\end{eqnarray}
where $\hat \gamma$ is the decoherence rate.
In this case, the master equation in Cartesian coordinates reads:
\ba
\dot {\bf{r}}= -2\hat\gamma(r_x, r_y, 0)^\top.
\label{pdmastereq}
\ea
For an initial quantum state with ${\bf{r}}_i:=(r_{ix}, r_{iy}, r_{iz})^\top$, 
the solution of the master equation (\ref{pdmastereq}) is given by:
\ba
{\bf{r}}(t)=(e^{-2\hat\gamma t }r_{ix}, e^{-2\hat\gamma t }r_{iy}, r_{iz})^\top.
\label{pdmastersol}
\ea
The locus of the fixed points for this model is given by the $z$-axis, i.e. it is the set of points with ${\bf{r}}_{\mathrm{fp}}=(0, 0, {\bar r}_{\mathrm{fp}})^\top$ and any ${\bar r}_{\mathrm{fp}}\in [0, 1]$, while the speed of purity change is:
\ba 
v_{\rm{PD}} (r, \theta)= -2\hat{\gamma} r^2 \sin^2\theta .
\ea
From Eq.~(\ref{pdmastersol}) and the trace distance condition (\ref{tracedcdt}) we then find that the 
relaxation time from $r_i$ to the fixed point in the absence of quantum control is:
\ba
T^{\mathrm{PD}}_{\mathrm{free}}({\bf{r}}_i; \epsilon)=\frac{1}{2\hat\gamma}\ln\left [\frac{\sqrt{r_{ix}^2+r_{iy}^2}}{\epsilon}\right ].
\label{pdtimefree}
\ea

In this case, since the locus of the fixed points is the whole $z$-axis, the task of speeding up the relaxation is ambiguous.
Given an arbitrary initial state ${\bf r}_i=(r_{ix},r_{iy},r_{iz})^\top$, the {\it natural} fixed point of the channel would be ${\bf r}_f=(0,0,r_{iz})^\top$.
Quantum control can then be used to achieve two different tasks: speeding up the relaxation  towards an arbitrary fixed point along the $z$-axis or, alternatively, towards the {\it natural} fixed point associated with the initial state.
The first task is trivial since it can be achieved instantaneously via a unitary rotation to the $z$-axis.
On the other hand, the second task is non-trivial and the optimal control strategy is analogous to the one used for the amplitude damping channel: one should first rotate the state to a position where the absolute value of the speed of purity change is maximum ({\it i.e.}\ to the equator), let the phase damping channel act and, once the desired purity is reached, perform a final rotation to the {\it natural} fixed point.
In this case, the corresponding optimal relaxation time is given (for $r_i > |r_{iz}| +\epsilon$) by:
\ba
T^{\mathrm{PD}}_{\mathrm{fast}}({\bf r}_i; \epsilon) =\frac{1}{2\hat\gamma}\ln\left [\frac{r_i}{|r_{iz}|+\epsilon}\right ].
\label{pdtimefast}
\ea

Comparing Eq. \rf{pdtimefree} with Eq. \rf{pdtimefast}, one can see that quantum control speeds up the relaxation for all initial states with $r_{iz} \ne 0$. However if we use, as done for the previous channels, 
the figure of merit based on the worst case scenario this advantage is lost. 
Indeed, it is easy to check that the maximum over ${\bf r}_i$ of the free evolution time~(\ref{pdtimefree}) and of the optimal relaxation time~(\ref{pdtimefast}) is, in both cases, equal to $ |\ln\epsilon|/(2\hat\gamma)$.

Furthermore, also for the phase damping channel, similarly to the case of the depolarizing channel, quantum control can keep the qubit near its initial state for indefinite time only if the initial state happens to be a fixed point along the z-axis.

\section{Discussion}
We have studied how the rate of relaxation of a qubit in the presence of some paradigmatic Markovian quantum channels (generalized amplitude damping, depolarization and phase damping) can be sped up or slowed down using optimal control. 
We analytically discussed the situation in which a generic initial state should reach the fixed point of the dynamics up to an arbitrarily small distance.
Our results suggest that optimal control cannot speed up the natural cooling rate of a thermal qubit in the presence of a cold bath.  However, it is possible to heat the qubit from an initial thermal state to its fixed point (another thermal state with lower purity) in finite time in the presence of a quantum control of large strength. We have also analyzed the relaxation of a qubit in the presence of a generic control Hamiltonian with infinitesimal strength $m$. Here the optimized relaxation time decreases linearly with $m$, with the slope depending on the explicit form of the Hamiltonian. We have also presented numerical data supporting our analytical results. 
Finally, we have given a measure of the performance of the quantum control in the worst case scenario, by maximizing the time duration of the evolutions with respect to the possible initial states of the qubit. Quantum control enhances this performance with respect to the uncontrolled decoherence in the cases of the generalized amplitude damping and depolarizing channels.
Time optimal control of a two-level dissipative quantum system has also been studied elsewhere \cite{sugny}-\cite{lapert13} using the Pontryagin maximum principle and geometrical methods \cite{jurdjevic}.  In our simplified approach, we further addressed the case of the time optimal relaxation of a qubit towards the fixed point of a depolarizing channel. Moreover, the inverse problem of slowing down the relaxation from an arbitrary initial quantum state of the qubit was not considered in \cite{sugny}-\cite{lapert13}. 
Note that this situation can be also thought as a "storage" procedure for certain special states.
We also found analytical expressions for the optimal time durations, which was possible in the geometric approach only for the saturation problem in NMR subject to longitudinal and transverse relaxation \cite{lapert11}. Finally, we considered the broader situation in which the final target of the quantum motion need not be reached exactly, but up to an arbitrarily small trace distance.
The next step would be to consider time optimal quantum control with fixed target fidelity for open systems in higher dimensions.

\section{Acknowledgements}
We thank D. Sugny for useful discussions and comments.
This work was supported by Regione Toscana, IP-SIQS, PRIN-MIUR, Progetto Giovani Ricercatori SNS and MIUR-FIRB-IDEAS project RBID08B3FM, SFB TR21 (CO.CO.MAT).

\section{appendix}

\subsection{Speed of change for the purity}

When we are only concerned about the quantum motion of the qubit along the radial coordinate, in other words when we are only interested in the speed of change of the purity of our quantum system, we have to study the quantity  $v=dP/dt$.
Using the relation $P=(1+r^2)/2$ and the master equation (\ref{parmaster}) in spherical coordinates, a simple algebra shows that the speed of change of the purity can be explicitly written in general as:
\ba
\frac{v(r, \theta, \varphi)}{r}&=&-(a_+-a_-)\cos\theta +2\Re [(d_+-d_-^\ast)e^{i\varphi}]\sin\theta 
\non \\
&+&\frac{r}{2}\{-(b +a_++a_-) +\Re(ce^{2i\varphi}) 
\non \\
&+&[b-a_+-a_--\Re(ce^{2i\varphi})]\cos 2\theta 
\non \\
&+&2\Re[(d_++d_-^\ast)e^{i\varphi}]
\sin 2\theta \},
\label{generalspeed}
\ea
where the coefficients $a_{\pm}, b, c, d_{\pm}$ depend upon the Lindblad operators in the following manner:
\ba
a_{\pm}&:=& \sum_a \gamma_a |l_{a \pm}|^2 ; ~~ b:= \sum_a \gamma_a (1+|l_{a z}|^2) \non\\
c&:=& \sum_a \gamma_a l_{a +}^\ast l_{a -} ; ~~ d_{\pm}:= \sum_a \gamma_a l_{a \pm}^\ast l_{a z},
\label{generalspeedcoef}
\ea
and we have defined $l_{a\pm} := l_{ax} \pm il_{ay}$.

\end{document}